\documentclass[useAMS,usenatbib]{mn2e}
\usepackage{wrapfig}
\usepackage{multicol}
\usepackage{epsfig}
\usepackage{multirow}

\title[Dispersal of molecular clouds by ionising radiation]
{Dispersal of molecular clouds by ionising radiation}
\author[S.~Walch et al.]
{S. Walch$^{1}$\thanks{E-mail: walch@mpa-garching.mpg.de}, 
A.P.~Whitworth$^{2}$, T.~Bisbas$^{3}$, R.~W\"unsch$^{4}$, D.~Hubber$^{5}$\\
$^{1}$Max-Planck-Institute for Astrophysics, Karl-Schwarzschild-Str. 1, 85741 Garching, Germany \\
$^{2}$School of Physics \& Astronomy, Cardiff University, 5 The Parade, Cardiff CF24 3AA, Wales, UK \\
$^{3}$UCL, Kathleen Lonsdale Building, Department of Physics \& Astronomy, Gower Place, London WC1E 6BT, UK\\
$^{4}$Astronomical Institute, Academy of Sciences of the Czech Republic, Bocni II 1401, 141 31 Prague, Czech Republic\\
$^{5}$Department of Physics and Astronomy, University of Sheffield, Hicks Building, Hounsfield Road, Sheffield, S3 7RH, UK}
\begin{document}

\date{Accepted . Received 2012 June 27; in original form }

\pagerange{\pageref{firstpage}--\pageref{lastpage}} \pubyear{2012}

\maketitle

\label{firstpage}

\begin{abstract}
The role of feedback from massive stars is believed to be a key element in the evolution of molecular clouds.
We use high-resolution 3D SPH simulations to explore the dynamical effects of a single O7 star emitting ionising photons at $10^{49}\,{\rm s}^{-1}$ and located at the centre of a molecular cloud with mass $10^4\,{\rm M}_{_\odot}$ and radius $6.4\,{\rm pc}$; we also perform comparison simulations in which the ionising star is removed. The initial internal structure of the cloud is characterised by its fractal dimension, which we vary between ${\cal D}=2.0$ and ${\cal D}=2.8$, and the roughly constant standard deviation, $\sigma_{_{10}} =0.38$, of its log-normal density PDF. (i) As regards star formation, in the short term ionising feedback is positive, in the sense that star formation occurs much more quickly (than in the control simulations), in gas that is compressed by the high pressure of the ionised gas. However, in the long term ionising feedback is negative, in the sense that most of the cloud is dispersed with an outflow rate of up to $\sim 10^{-2} {\rm M}_\odot {\rm yr}^{-1}$, on a timescale comparable with the sound-crossing time for the ionised gas ($\sim 1\;{\rm to}\;2\,{\rm Myr}$), and triggered star formation is therefore limited to a few percent of the cloud's mass. We will describe in greater detail the statistics of the triggered star formation in a companion paper. (ii) As regards the morphology of the ionisation fronts (IFs) bounding the H{\sc ii} region and the systematics of outflowing gas, we distinguish two regimes. For low ${\cal D}\la 2.2$, the initial cloud is dominated by large-scale structures, so the neutral gas tends to be swept up into a few extended coherent shells, and the ionised gas blows out through a few large holes between these shells; we term these H{\sc ii} regions {\it shell-dominated}. Conversely, for high ${\cal D}\ga 2.6$, the initial cloud is dominated by small-scale structures, and these are quickly overrun by the advancing IF, thereby producing neutral pillars protruding into the H{\sc ii} region, whilst the ionised gas blows out through a large number of small holes between the pillars; we term these H{\sc ii} regions {\it pillar-dominated}. (iii) As regards the injection of bulk kinetic energy, by $\sim 1\,{\rm Myr}$, the expansion of the H{\sc ii} has delivered a mass-weighted root-mean-square velocity of $\sim 6\,{\rm km}\,{\rm s}^{-1}$; this represents less than $0.1\%$ of the total energy radiated by the O7 star.
\end{abstract}


\section{Introduction}

Infrared shells and bubbles \citep{Churchwell2006, Churchwell2008} are ubiquitous in the Galaxy and are often associated with H{\sc ii} regions, i.e. localised regions in which the Lyman continuum radiation ($E_\gamma >13.6\,{\rm eV}$) from one or more young, massive stars ionises the gas and heats it to $T_{_{\rm I}}\sim 10^4\,{\rm K}$. Because the neutral gas outside the H{\sc ii} region is much colder ($T_{_{\rm N}}\sim 10\;{\rm to}\;30\,{\rm K}$), the H{\sc ii} region is overpressured, and expands, sweeping up the neutral gas. Molecular clouds typically have a complicated clumpy structure, and so the ionising photons can penetrate to different distances in different directions, causing highly irregular ionisation fronts \citep[e.g.][]{Hegmann2003}. Consequently, large-scale H{\sc ii} regions have complex shapes and diverse morphologies. Sometimes they appear as bright-rimmed clouds \citep{Urquhart2009} or perfectly round shells \citep{Deharveng2009, Deharveng2010}, other times they are dominated by pillar-like structures \citep{Sugitani2002, Smith2010, Preibisch2011}, or a mixture of all three features \citep{Koenig2008}. These diverse morphologies must encode important clues as to how a massive star interacts with the surrounding interstellar medium (ISM).

Feedback from young, massive stars plays a critical role in the evolution of molecular clouds, and of the galaxies in which they reside. The input of radiative and mechanical energy due to ionising radiation and stellar winds from massive stars is most influential during the early evolution of a molecular cloud, before the first supernova explodes after $\ga 3\,{\rm Myr}$. In particular, the expansion of H{\sc ii} regions may have a positive effect and trigger new star formation \citep{Elmegreen1998, Elmegreen2002b, Whitworth2002, Elmegreen2011}, but it may also have a negative effect by efficiently dispersing the surrounding cloud \citep{Whitworth1979, Tenorio1979, Bodenheimer1979, Williams1997, Franco1990, Matzner2002}, thereby terminating star formation, and possibly also unbinding the newly-formed cluster of stars \citep{Dale2005}. Ionising feedback may also play a key role in regulating the star formation efficiency on galactic scales \citep{McKee1989, Vazquez2010}; and H{\sc ii} regions and bubbles excavated by ionisation may be the first step towards launching large scale winds and outflows from galactic disks \citep{Tenorio2003, Hopkins2012}.

Many numerical studies concerned with positive feedback, i.e. the triggering of star formation, have been performed in recent years: on individual molecular cloud cores \citep{Bisbas2011, Haworth2012}; in the context of bright rims and pillars  \citep{Elmegreen1995, Miao2006, Gritschneder2009, Gritschneder2010, Ercolano2011, Mackey2011}, on parsec-scale molecular clouds \citep{Mellema2006, Krumholz2007, Arthur2011, Walch2011}; and on whole star clusters \citep{Dale2011b}. Observationally, it is very difficult to distinguish triggered from spontaneous (untriggered) star formation, even though many authors have tried to identify age gradients in the young stars surrounding H{\sc ii} regions \citep{Preibisch2007}, and to establish evidence for the {\it Collect \& Collapse} scenario in swept-up shells \citep{Zavagno2006, Deharveng2008, Zavagno2010}. 

The negative effect of ionising feedback, i.e. its role is dispersing molecular clouds, has been evaluated analytically by \citet{Whitworth1979}. \citet{Matzner2002} has considered cloud destruction by all stellar feedback mechanisms and concludes that ionising feedback will destroy normal giant molecular clouds in less than a crossing time. The effect of ionsation feedback has been explored numericlly by \citet{Dale2005}. Based on simulations by \citet{Bonnell2002}, they model the impact of ionising feedback in turbulent, star-cluster-forming clouds, and find that photoionisation is able to unbind star clusters by efficiently removing gas quite early on during their evolution. They note that the mean gas density of the affected cloud is the critical parameter. \citet{Krumholz2006} study the disruptive power of H{\sc ii} regions using semi-analytical models, and conclude that lower mass clouds ($M\la 3\times 10^5\,{\rm M}_\odot$) may be destroyed in $\sim 10\,{\rm Myr}$. In more massive clouds disruption by ionisation plays a less important role \citep{Krumholz2009, Murray2010, Fall2010}. These conclusions have recently been confirmed in 3D SPH simulations by \citet{Dale2012}, who study star-cluster forming clouds with $10^4\;{\rm to}\;10^6\,{\rm M}_\odot$; their lower mass clouds are comparable with the ones we investigate in the current study. They find that massive clouds resist dispersal by ionising feedback, because it is quenched by accretion flows. Quenching of feedback by strong accretion is also important in the early stages of H{\sc ii} region formation, i.e. in the ultra-compact stage \citep{MacLow2007,Peters2010, Peters2011}. Moreover, the location of a massive star is important for the dispersal efficiency. This has first been noted by \citet{Mazurek1980} and \citet{Yorke1989}, and recently confirmed by \citet{Gendelev2012}, who perform three-dimensional radiation-magneto-hydrodynamic simulations of blister-type H{\sc ii} regions. They note that the recoil provided by escaping gas may also play a role in destroying the parental molecular cloud.

Feedback also drives turbulent motions in molecular clouds. \citet{Matzner2002} argues that H{\sc ii} regions are the main drivers of molecular cloud turbulence, being energetically dominant over the combined effects of stellar winds and supernovae. \citet{Mellema2006} record that the level of molecular cloud turbulence is maintained at $\sim 8\,{\rm km}\,{\rm s}^{-1}$ in their simulations of ionising feedback, but they also note that radial velocities account for a significant fraction of the total kinetic energy. \citet{Gritschneder2009} discuss turbulence driving in their simulations of ionising feedback, and report that the energetic ratio of solenoidal to compressible modes \citep{Federrath2008} is significantly reduced, relative to the fiducial ratio of 2:1.

In this paper we simulate the development of large-scale H{\sc ii} regions in molecular clouds having different initial fractal dimension $\mathcal{D}$. In contrast with \citet{Dale2005} our molecular clouds all have the same mass, radius and density PDF. We demonstrate that the full range of morphological features observed in H{\sc ii} regions might be attributable to the fractal dimension, ${\cal D}$, of the parental molecular cloud. As ${\cal D}$ is increased from small values ${\cal D}\sim 2$, there is a shift from H{\sc ii} regions bounded by relatively smooth extended shell-like IFs, with ionised gas blowing out through large holes between these shells, to H{\sc ii} regions bounded by much more irregular IFs with numerous pillars protruding into the H{\sc ii} region, and ionised gas blowing out through many small holes. The kinetic energy injected into the ISM is essentially independent of ${\cal D}$, and the time taken to disperse the cloud is short ($\sim 1\,{\rm Myr}$ for ${\cal D}=2.8\;$ to $\;\sim 2\,{\rm Myr}$ for ${\cal D}=2.0$).

The plan of the paper is the following. In section \ref{sec2} we describe the generation of fractal molecular clouds and the numerical method, including the treatment of the ionising radiation. In section \ref{sec3} we analyse the large-scale structure of the H{\sc ii} regions obtained with different fractal dimensions, and in section \ref{sec4} we analyse the structure of the corresponding IFs and the associated outflow rates. In section \ref{sec5} we evaluate the injection of kinetic energy into the ISM. In section \ref{sec6} we summarise our main conclusions.

\section{Initial conditions and numerical method}\label{sec2}%

\subsection{Generation of fractal molecular clouds with an FFT algorithm}\label{fractal}

It is well known that molecular clouds are rich in internal structure, which may derive from pure turbulence \citep[e.g.][]{Klessen2001, Padoan2002}, or from gravitational amplification of noise in the density field \citep{Bonnell2002} or from initial turbulence leading to density fluctuations which are subsequently enhanced by gravity \citep{Heitsch2009, Ballesteros2011a, Ballesteros2011b, Walch2010, Walch2012, Girichidis2011, Girichidis2012}. It is unclear which of these processes is dominant in shaping molecular clouds. However, what we do know is that on scales from the typical prestellar core size of $\sim 0.1\,{\rm pc}$, up to $\sim 100\,{\rm pc}$ \citep{Bergin2007}, molecular clouds subscribe to a statistically self-similar fractal structure \citep{Stutzki1998, Elmegreen1996, Falgarone1991}. 

Apart from their applicability, we adopt fractal initial conditions for their practicality. Fractals with mass-size relations similar to those observed \citep{Larson1981} are easily created in Fourier space \citep{Elmegreen2002, Shadmehri2011}, whilst at the same time retaining the freedom to choose the fractal dimension, ${\cal D}$, arbitrarily. In contrast, when starting with a periodic box of driven or decaying turbulence \citep[see e.g.][for particularly nice simulations of this type]{Arthur2011}, $\mathcal{D}$ is a consequence of the physics implemented, and cannot easily be adjusted.

\citet{Stutzki1998} have shown that a three-dimensional density field in which the power spectrum of the density fluctuations has index $-n$ (i.e. $P(k)\propto k^{-n}$) has box-coverage fractal dimension
\begin{equation}
\mathcal{D}= 4\,-\,\frac{n}{2}
\end{equation}  
\citep[see also][]{Federrath2009}. Thus, defining $\mathcal{D}$ is equivalent to defining the power spectral index $n$, and vice versa. Low ${\cal D}$ corresponds to high $n$, i.e. a density field dominated by a small number of extended density structures; conversely, high ${\cal D}$ corresponds to low $n$, i.e. a density field dominated by a large number of compact density structures.

CO line surveys are well suited to determining the structures of molecular clouds like the ones we simulate here, because this transition is a dominant coolant at number densities $\sim 10^2 - 10^3\,{\rm cm}^{-3}$. Values of the fractal dimension estimated in this way are typically $\mathcal{D}\approx 2.3 \pm 0.3$ \citep{Elmegreen1996}. For example, observations of molecular clouds in the Milky Way typically yield $\mathcal{D}\approx 2.4$ \citep{Falgarone1991, Vogelaar1994, Stutzki1998, Lee2004, Sanchez2005}, corresponding to $n = 3.2$. For Ophiuchus, Perseus, and Orion, \citet{Sanchez2007} obtain a slightly higher value, $\mathcal{D}=2.6\pm 0.1$ ($n=2.8\pm 0.2$), and \citet{Miville2010} derive $\mathcal{D}\approx2.65\pm0.05$ ($n\approx 2.7\pm 0.1$) from Herschel-SPIRE observations of the Polaris Flare. \citet{Sanchez2007} and \citet{Schneider2011} report differences between low-mass star forming regions and massive, giant molecular clouds, based on the $\Delta-$variance method applied to $A_V$ maps \citep{Ossenkopf2008}, where $A_V$ is the visual extinction. In M33, \citet{Sanchez2010} find $\mathcal{D}\simeq 2.2\;{\rm to}\;2.5$ on scales $\la 500\,{\rm pc}$, and higher values of $\mathcal{D}$ on larger spatial scales.

Here, the initial three-dimensional fractal density field is constructed in a periodic box, using an FFT-based algorithm \citep{Shadmehri2011}. The algorithm has three main input parameters: (i) the power spectral index, $n=2(4-\mathcal{D})$, (ii) the random seed ${\cal R}$ used to generate different realisations; and (iii) the density scaling constant $\rho_0$ (see Eqn. \ref{EQN:DENSITYFIELD} below). We populate the integer modes $k=1,..,128$ along each Cartesian axis $(x, y, z)$, where $k=1$ corresponds to the linear size of the box ($15\,{\rm pc}$). After generating the spectrum of density fluctuations in Fourier space, and transforming them to obtain $\rho_{_{\rm FFT}}(x,y,z)$ in real space, this field is scaled according to
\begin{equation}\label{EQN:DENSITYFIELD}
\rho(x,y,z)=\exp\left\{\frac{\rho_{_{\rm FFT}}(x,y,z)}{\rho_0} \right\}\,. \label{EqScale}
\end{equation}
The resulting density field, $\rho(x,y,z)$, has a log-normal density PDF and a clump mass distribution in agreement with observations \citep[as described in][]{Shadmehri2011}. For a given spectrum of density fluctuations, changing $\rho_0$ changes the standard deviation, $\sigma$, of the log-normal density PDF, whilst leaving the underlying topology of the density field unchanged: small $\rho_{_{\rm O}}$ produces a very wide density PDF (large $\sigma$), and conversely large $\rho_{_{\rm O}}$ produces a very narrow density PDF (small $\sigma$). Before creating the initial SPH particle distribution, we shift the point of maximum density to the centre of the computational domain and cut out a sphere of diameter $12.8\,{\rm pc}$ centred on this point. Finally, the box is partitioned with a $128^3$ grid, which is then populated with SPH particles, by placing the appropriate number of particles randomly within each grid cell. We have checked the fidelity of this procedure by performing a back transformation and recovering the correct power spectrum.

\subsection{Numerical method}

We use the SPH code \textsc{Seren} \citep{Hubber2011}, which is well-tested and has already been applied to many problems in star formation \citep[e.g.][]{Walch2011, Bisbas2011, Stamatellos2011}. We employ the standard SPH implementation \citep{Monaghan1992}. The SPH equations of motion are solved with a second-order leapfrog integrator, in conjunction with an hierarchical, block time-stepping scheme. Gravitational forces are calculated using an octal spatial decomposition tree \citep{Barnes1986}, with monopole and quadrupole terms and a Gadget-style opening-angle criterion \citep{Springel2001}. We use the standard artificial viscosity prescription \citep{Monaghan1983}, moderated with a Balsara switch \citep{Balsara1995}. Ionising radiation is treated with an HEALPix-based adaptive ray-splitting algorithm, which allows for optimal resolution of the ionisation front in high resolution simulations \citep[see][]{Bisbas2009}. We assume the On-The-Spot approximation and do not treat the diffusive radiation field.

The temperature of ionised gas particles is set to $T_{_{\rm I}}=10^4\,{\rm K}$. The temperature of neutral gas is given by a barotropic equation of state,
\begin{equation}
  T(\rho)=T_{_{\rm N}}\left\{1+\left(\frac{\rho}{\rho_{_{\rm CRIT}}}\right)^{(\gamma-1)}\right\}\,,
\end{equation}
where $T_{_{\rm N}}=30\,\rm{K}$, $\rho_{_{\rm CRIT}}=10^{-13}\,{\rm g}\,{\rm cm}^{-3}$, and $\gamma=5/3$. The choice of $T_{_{\rm N}}=30\,{\rm K}$ influences the gravitational stability of the neutral gas swept up by the expanding H{\sc ii} region. If the shell were able to cool to lower temperature, it would fragment more readily than currently seen in our simulations. For this reason we will explore a more realistic cooling function in a future paper. With the above prescriptions for the temperature, and assuming solar elemental composition, the isothermal sound speed is $c_{_{\rm I}}=12\,{\rm km}\,{\rm s}^{-1}$ in ionised gas, and $c_{_{\rm N}}=0.32\,{\rm km}\,{\rm s}^{-1}$ in neutral gas at low density ($\rho\ll\rho_{_{\rm CRIT}}$).

Although we are not concerned with triggered star formation in this paper, we introduce sinks at density peaks above $\rho_{_{\rm SINK}}=10^{-11}\,\rm{g}\,\rm{cm}^{-3}$, provided that the density peak in question is at the bottom of its local gravitational potential well \citep[see ][in preparation]{Hubber2011inprep}. Since $\rho_{_{\rm SINK}}\!\gg\!\rho_{_{\rm CRIT}}$, a condensation that is converted into a sink is always well into its Kelvin-Helmholtz contraction phase. Once formed, a sink is able to accrete gas smoothly from its surroundings and thereby grow in mass.

\begin{table}
\begin{center}
\begin{tabular}{ccccc}
\hline\hline
ID$\;\;$ & $\mathcal{D}$ & $10^{21}{\bar\rho}_M$ & $\sigma_{_{10}}$ & $t_1,\,t_{15}$ \\
& & $\overline{{\rm g}\,{\rm cm}^{-3}}$ & & $\overline{\rm Myr}$ \\
\hline
${\cal D}2.0/{\rm O}7$ & 2.0 & 1.29 & 0.39 & 0.51,\,0.77 \\
${\cal D}2.2/{\rm O}7$ & 2.2 & 1.17 & 0.39 & 0.48,\,0.65 \\
${\cal D}2.4/{\rm O}7$ & 2.4 & 1.02 & 0.38 & 0.44,\,0.59 \\
${\cal D}2.6/{\rm O}7$ & 2.6 & 0.98 & 0.38 & 0.49,\,0.77 \\
${\cal D}2.8/{\rm O}7$ & 2.8 & 0.93 & 0.37 & 0.62,\,0.89 \\
${\cal D}2.0/{\rm NI}$ & 2.0 & 1.29 & 0.39 & 0.98,\,0.99 \\
${\cal D}2.2/{\rm NI}$ & 2.2 & 1.17 & 0.39 & 0.94,\,0.99 \\
${\cal D}2.4/{\rm NI}$ & 2.4 & 1.02 & 0.38 & $>\!1.0,\,>\!1.0$ \\
${\cal D}2.6/{\rm NI}$ & 2.6 & 0.98 & 0.38 & $>\!1.0,\,>\!1.0$ \\
${\cal D}2.8/{\rm NI}$ & 2.8 & 0.93 & 0.37 & $>\!1.0,\,>\!1.0$ \\
\hline
\end{tabular}
\caption{Initial conditions. Column 1 gives the run ID (where the first element of the ID gives the fractal dimension, and the second records whether there is an ionising star, O7, or no ionising star, NI), column 2 the fractal dimension, ${\cal D}$, column 3 the mass-weighted mean density, ${\bar\rho}_M$, and column 4 the logarithmic standard deviation, $\sigma_{_{10}}$, of the mass-weighted density PDF (see Fig. \ref{FIG:RHOPDF_PDF_ini}). Column 5 gives $t_1$ and $t_{15}$, where $t_1$ is the time at which the first sink forms, and $t_{15}$ is the time at which the fifteenth sink forms.}
\label{Table 1}
\end{center}
\end{table}

\subsection{Initial conditions}

\begin{figure}
\begin{center}
\includegraphics[width=90mm, angle=0]{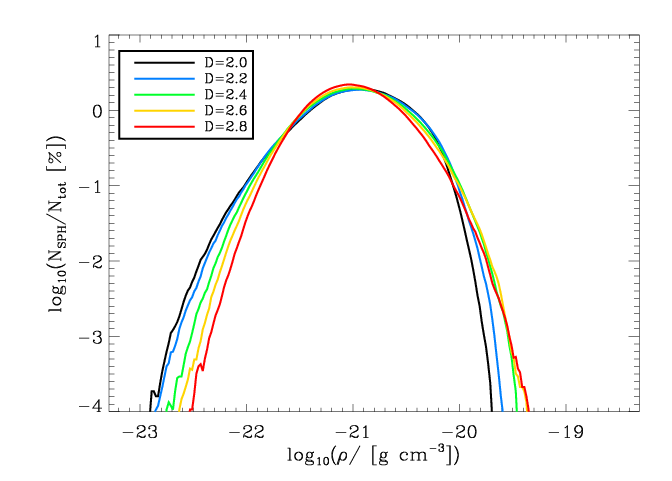} 
\caption{Initial mass-weighted density PDFs for ${\cal D}=2.0,\,2.2,\,2.4,\,2.6\;{\rm and}\;2.8$. ${\rm N}_{_{\rm SPH}}$ is the number of SPH particles in each density bin and ${\rm N}_{_{\rm TOT}}$ is the total particle number. Means and standard deviations are given in Table \ref{Table 1}.}
\label{FIG:RHOPDF_PDF_ini}
\end{center}
\end{figure}

We consider a single cloud mass, $10^4\,{\rm M}_\odot$, and a single cloud radius, $6.4\,\rm{pc}$, so the volume-weighted mean density of the cloud is $\bar{\rho}_V= 0.62\times 10^{-21}\,{\rm g}\,{\rm cm}^{-3}$, the escape velocity from the surface of the cloud is $v_{_{\rm ESC}}= 3.7\,{\rm km}\,{\rm s}^{-1}$, and the mean freefall time is ${\bar t}_{_{\rm FF}}\sim 3\,{\rm Myr}$. Every cloud is modelled with a total number of ${\rm N}_{_{\rm TOT}}=2.5 \times 10^6$ SPH particles, resulting in a mass resolution of $0.4 M_\odot$ \citep{BateBurkert97}.The only parameter we vary is the initial fractal dimension, ${\cal D}=2.0,\,2.2,\,2.4,\,2.6\;{\rm and}\;2.8$. For each value of ${\cal D}$, we generate a single realisation, using the same random seed, ${\cal R}\!=\!1$, so that the density fields in the five different clouds have the same underlying topology. We also adjust $\rho_0$ so that the (approximately log-normal) density PDFs of the five clouds have roughly the same standard deviation (see Fig.\ref{FIG:RHOPDF_PDF_ini} for the initial mass-weighted density PDFs). We then simulate the evolution of each cloud twice, once with a single O7 star at the centre of the cloud, emitting ionising photons at $\dot{\cal N}_{_{\rm LyC}}=10^{49}\,{\rm s}^{-1}$, and once with no ionising star. Table \ref{Table 1} lists the ID of each simulation, the fractal dimension, the mass-weighted mean density, $\bar{\rho}_M$, the standard deviation of the logarithmic density PDF, $\sigma_{_{10}}$, and the times, $t_{_1}$ and $t_{_{15}}$, at which the first and fifteenth sinks are created (a measure of how rapidly star formation occurs). We note that $\bar{\rho}_M$ is inevitably larger (up to a factor of 2) than $\bar{\rho}_V$. In the lefthand column of Fig. \ref{FIG_M1} we show initial column-density images of all five clouds, as seen projected on the (x,y)-plane. We note that as ${\cal D}$ is increased, the scale of the dominant features in these images becomes smaller. In addition to these ten simulations involving lumpy clouds, we perform a control simulation with a uniform-density cloud of the same mass and radius: this has the ID `{\sc control}'.

\section{General evolution}\label{sec3}%

\subsection{Morphology}

\begin{figure*}
\includegraphics[width=195mm]{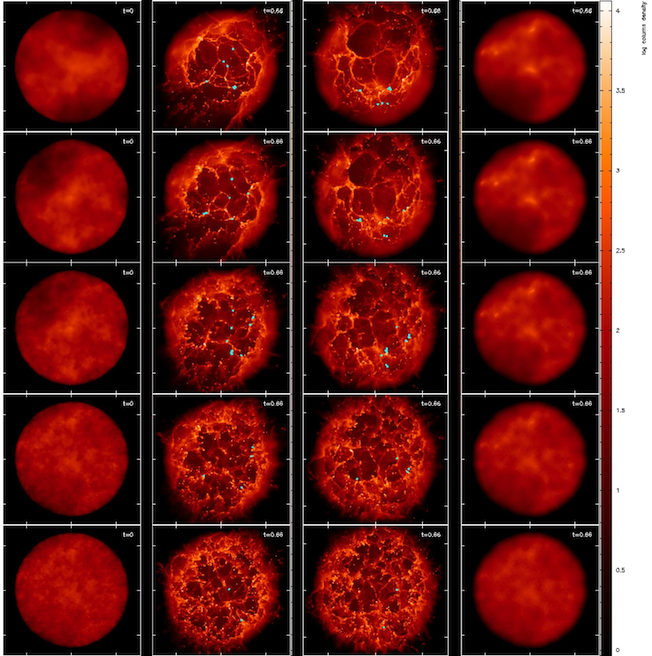}
\caption{From top to bottom, the different rows correspond to $\mathcal{D}=2.0,\,2.2,\,2.4,\,2.6\;{\rm and}\;2.8$. The lefthand column shows column-density images of the initial fractal clouds, projected onto the ($x,y$)-plane. The middle two columns show column-density images at $t=0.66\,{\rm Myr}$ from the simulations with ionising radiation, projected onto the ($x,y$)- and ($x,z$)-planes. The righthand column shows column-density images at $t=0.66\,{\rm Myr}$ from the simulations without ionising radiation. The colour-table on the right gives the logarithm of column-density in ${\rm M}_\odot\,{\rm pc}^{-2}$.}
\label{FIG_M1}
\end{figure*}

The morphology of the evolving H{\sc ii} region is strongly dependent on the fractal dimension of the initial molecular cloud. In the two central columns of Fig. \ref{FIG_M1} we show column-density images of the five clouds as projected on the (x,y)- and (x,z)-planes, at time $t=0.66\,{\rm Myr}$. By this time there has been considerable interaction between the H{\sc ii} region and the molecular gas, and in all cases star formation has been triggered (turquoise dots mark sink particles); we will analyse the triggering of star formation in a companion paper (Walch et al., in prep.).

Although the simulated H{\sc ii} regions have highly chaotic structures (which could be explored further by invoking different random seeds, ${\cal R}$), there are also discernable statistical trends with changing ${\cal D}$. For low ${\cal D}$, the dominant density structures in the initial cloud have large coherence lengths, and consequently the H{\sc ii} regions tend to be bounded by extended, relatively smooth shells, and their external appearance can be strongly dependent on viewing angle. Between the shells are large holes, through which the ionised gas can escape into the ISM. We term these H{\sc ii} regions {\it shell-dominated}. Conversely, for large ${\cal D}$, the dominant density structures in the initial cloud have relatively short coherence lengths, and consequently the H{\sc ii} regions tend to have a much more irregular boundary, with many neutral pillars protruding into the ionised gas. There are many small holes through which the ionised gas escapes. We term these H{\sc ii} regions {\it pillar-dominated}. The transition between shell-dominated and pillar-dominated H{\sc ii} regions occurs at ${\cal D}\!\sim\!2.4$, which is the average fractal dimension of many observed molecular clouds (see section \ref{fractal}). 

In the righthand column of Fig. \ref{FIG_M1} we show, for comparison, column-density images projected on the (x,y)-plane for the simulations performed without an ionising star. Under this circumstance the evolution proceeds on a free-fall timescale, $\sim 3\,{\rm Myr}$, so that after $0.66\,{\rm Myr}$ there has been much less change to the density field.

We conclude that the main requirement for the formation of pillars is the existence of small-scale structure in the density field into which the IF advances. As ${\cal D}$ increases, and therefore $n$ decreases, there is more small-scale structure and more pillars are formed. Since, for turbulent gas with a fixed ratio of solenoidal to compressive modes, small-scale structure should become more abundant with increasing Mach Number, this conclusion agrees with the findings of \citet{Gritschneder2010}. They have performed simulations of the ionisation of a slab of turbulent interstellar gas, and invoke different Mach numbers (where the Mach Number is measured relative to the sound speed in the neutral gas). They conclude (i) that the most favourable regime for the formation of pillar-like structures is Mach 4 to 10, and (ii) that the typical size of a pillar is set by the scale of the largest turbulent perturbation. 

\subsection{Radius of the H{\sc ii} region}

\begin{figure}
\includegraphics[width=80mm]{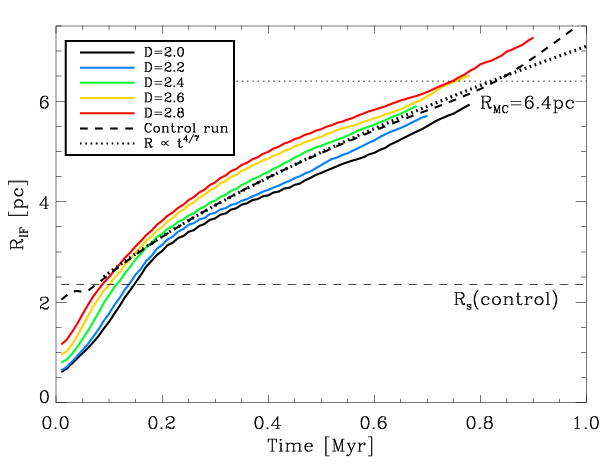} 
\caption{The solid curves give the mean radius of the ionisation front, $\bar{R}_{_{\rm IF}}$, as a function of time, for different $\mathcal{D}$ (see key for colour coding of different ${\cal D}$). The dashed curve represents the uniform-density {\sc control} simulation, and the dotted curve shows $\bar{R}_{_{\rm IF}}\propto t^{4/7}$. The thin, dashed horizontal line marks the Str{\o}mgren radius of the {\sc control} run, and the thin, dotted horizontal line marks the initial cloud radius.}
\label{FIG_IFPOS}
\end{figure}

The mean radius of the ionisation front, ${\bar R}_{_{\rm IF}}$ (see Fig. \ref{FIG_IFPOS}), is an observational quantity which is easily obtained \citep{Churchwell2006}, and frequently used in statistical studies \citep[e.g. in studies of triggered star formation by][]{Thompson2011}. Fig. \ref{FIG_IFPOS} shows the evolution of ${\bar R}_{_{\rm IF}}$ obtained with different ${\cal D}$. $\bar{R}_{_{\rm IF}}$ is estimated by averaging the radii of all SPH particles having temperature between $500\,{\rm K}$ and $5000\,{\rm K}$ \citep[see][]{Bisbas2009}.

Given a uniform-density ambient medium, as in the {\sc control} simulation, the IF advances to the Str{\o}mgren radius,
\begin{equation}
R_{_{\rm S}} = \left(\frac{3\,\dot{\cal N}_{_{\rm LyC}}\,m_{_{\rm H}}^2}{4\,\pi\,\alpha_{_{\rm B}}\,X^2\,\rho^2}\right)^{1/3}\,,
\end{equation}
in a few recombination times,
\begin{eqnarray}
t_{_{\rm REC}}&=&\frac{m_{_{\rm H}}}{\alpha_{_{\rm B}}\,X\,\bar{\rho}_V}\;\,=\;\,0.4\,{\rm kyr}.
\end{eqnarray}
Here 
$m_{_{\rm H}}=1.7\times 10^{-24}\,{\rm g}$ is the mass of an hydrogen atom, $\alpha_{_{\rm B}}=2.6\times 10^{-13}\,{\rm cm}^3\,{\rm s}^{-1}$ is the Case B recombination coefficient (i.e. invoking the On-The-Spot Approximation, and therefore neglecting recombinations straight into the ground state), and $X=0.7$ is the fraction by mass of hydrogen. For the {\sc control} simulation we derive $R_{_{\rm S}}=2.3\,{\rm pc}$. Thereafter the IF quickly switches to D-type, and the radius is given approximately by \citep{Shu1991}
\begin{eqnarray}
\frac{R_{_{\rm IF}}(t)}{R_{_{\rm S}}}&\simeq&\left\{1+\frac{7\,c_{_{\rm I}}\,t}{4\,R_{_{\rm S}}}\right\}^{4/7}\,.
\end{eqnarray}
This solution is very well reproduced in the {\sc control} simulation, until the IF reaches the edge of the cloud (see dashed and dotted lines on Fig. \ref{FIG_IFPOS}).

For the fractal clouds, ${\bar R}_{_{\rm IF}}$ is initially smaller than $R_{_{\rm S}}$, because the O7 star is placed at the densest point in the initial density field. However, as the IF advances away from this point, the density falls below $\bar{\rho}_V$, so that $\bar{R}_{_{\rm IF}}$ expands faster than $\propto t^{4/7}$ and catches up with the uniform-density solution after $\sim 0.2\,{\rm Myr}$. Thereafter $\bar{R}_{_{\rm IF}}$ tracks the uniform-density solution quite closely ($\pm 0.5\,{\rm pc}$) for all ${\cal D}$. For fractal clouds, there are two competing effects: (i) $\bar{R}_{_{\rm IF}}$ is reduced because the volume-weighted mean square density, $\left<\rho^2\right>_V$, is higher than for the uniform density in the {\sc control} simulation, and therefore the ionising photons should get used up balancing recombination in a smaller volume; (ii) $\bar{R}_{_{\rm IF}}$ is increased because the IF can break out of the cloud more readily, through paths of low emission measure. Effect (i) appears to dominate for low ${\cal D}$, making $\bar{R}_{_{\rm IF}}$ smaller; and effect (ii) appears to dominate for high ${\cal D}$, making $\bar{R}_{_{\rm IF}}$ larger.

\section{Structural analysis of shells}\label{sec4}%

From Fig. \ref{FIG_M1}, it is evident that, over and above stochastic variations, there is a systematic progression in the structural properties of the H{\sc ii} regions with increasing ${\cal D}$. In this section, we analyse the shell structures more quantitatively, and compute outflow rates of hot and cold gas, in order to derive cloud destruction timescales. 

\subsection{Shell structure}

\begin{figure}
\begin{center}
\includegraphics[width=78mm]{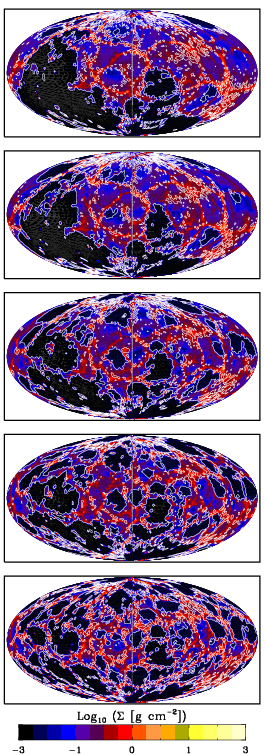} \\
\caption{Hammer projections of the column-density, $\Sigma(\theta,\phi)$, as seen from the ionising star at time $t\!=\!0.66\,{\rm Myr}$. From top to bottom, $\mathcal{D}=2.0,\,2.2,\,2.4,\,2.6\;{\rm and}\;2.8$. The white contours mark the separatrices between ionisation-bounded directions and density-bounded directions.}
\label{FIG:HAMMER}
\end{center}
\end{figure}

Fig. \ref{FIG:HAMMER} shows Hammer projections of the column-density, as seen from the ionising star, at $t=0.66\,{\rm Myr}$. If we adopt spherical polar coordinates, ($r,\theta,\phi$), centred on the ionising star, then the false colour on the Hammer projection represents the column-density, $\Sigma$, as seen in each direction, $(\theta,\phi)$, from the ionising star,
\begin{eqnarray}
\Sigma(\theta,\phi)&=&\int\limits_{r=0}^{r=\infty}\rho(r,\theta,\phi)\,dr\,.
\end{eqnarray}
For low ${\cal D}$, the Hammer projection is dominated by a small number of extended column-density features, which  correspond to the extended convex shell-like segments of the ionisation front seen in the top rows of Fig. \ref{FIG_M1}; because there is only a small number of extended features these H{\sc ii} regions can look quite different to different external observers (compare the middle images on the top two rows of \ref{FIG_M1}). Conversely, for high ${\cal D}$, the Hammer projection is dominated by a larger number of more compact column-density features, corresponding to small concave bright-rimmed clouds and pillars, as seen in the bottom rows of Fig. \ref{FIG_M1}; because there is a large number of compact features, these H{\sc ii} regions look rather similar to different external observers (compare the middle images on the bottom two rows of Fig. \ref{FIG_M1}).

The white contours on Fig. \ref{FIG_M1} demark the separatrices between directions in which the H{\sc ii} region is ionisation-bounded (i.e. rays along which the ionising photons run out before reaching the edge of the cloud), and directions in which the H{\sc ii} region is density-bounded (i.e. rays along which there is still ionising radiation left at the edge of the computational domain). Table \ref{TAB:fIB} gives the fraction of the sky, seen from the ionising star, that is ionisation bounded,
\begin{eqnarray}
f_{_{\rm IB}}&=&\frac{1}{4\pi}\,\int_{_{\rm IB}}\,d\Omega\,,
\end{eqnarray}
for ${\cal D}=2.0\;{\rm and}\;2.8$, at times $t=0.30\;{\rm and}\;0.66\,{\rm Myr}$. The large-scale holes that dominate the low ${\cal D}$ cases form relatively quickly after the ionising star switches on, as their formation is based on the pre-existence of coherent regions of low density which are too extended to be filled in significantly by the erosive flows at their boundaries. However, because the net length of these boundaries is relatively short, the net rate at which the remaining neutral gas is eroded is slow. Consequently $f_{_{\rm IB}}$ starts small but decreases slowly in these cases. In contrast, the small-scale holes that dominate the high ${\cal D}$ cases are initially choked because they are small and therefore the erosive flows at their boundaries are much more effective at inhibiting the escape of ionising radiation. However, because the net length of these boundaries is relatively large, the net rate of erosion is high. Consequently $f_{_{\rm IB}}$ starts high, but decreases rapidly in these cases.

\begin{table}
\begin{center}
\begin{tabular}{ccc}
\hline\hline
$\,\hspace{0.8cm}\,$ & $\hspace{0.3cm}{\cal D}=2.0\hspace{0.3cm}$ & $\hspace{0.3cm}{\cal D}=2.8\hspace{0.3cm}$ \\
\hline
$t\!=\!0.30\,{\rm Myr}$ & $f_{_{\rm IB}}\!=\!0.72$ & $f_{_{\rm IB}}\!=\!0.80$ \\
$t\!=\!0.66\,{\rm Myr}$ & $f_{_{\rm IB}}\!=\!0.62$ & $f_{_{\rm IB}}\!=\!0.45$ \\
\hline
\end{tabular}
\caption{The fraction of the sky that is ionisation bounded, as seen from the ionising star, $f_{_{\rm IB}}$, for ${\cal D}=2.0\;{\rm and}\;2.8$, at times $t=0.30\;{\rm and}\;0.66\,{\rm Myr}$.}
\label{TAB:fIB}
\end{center}
\end{table}

\subsection{Outflow rates}

\begin{figure}
\begin{center}
\includegraphics[width=90mm]{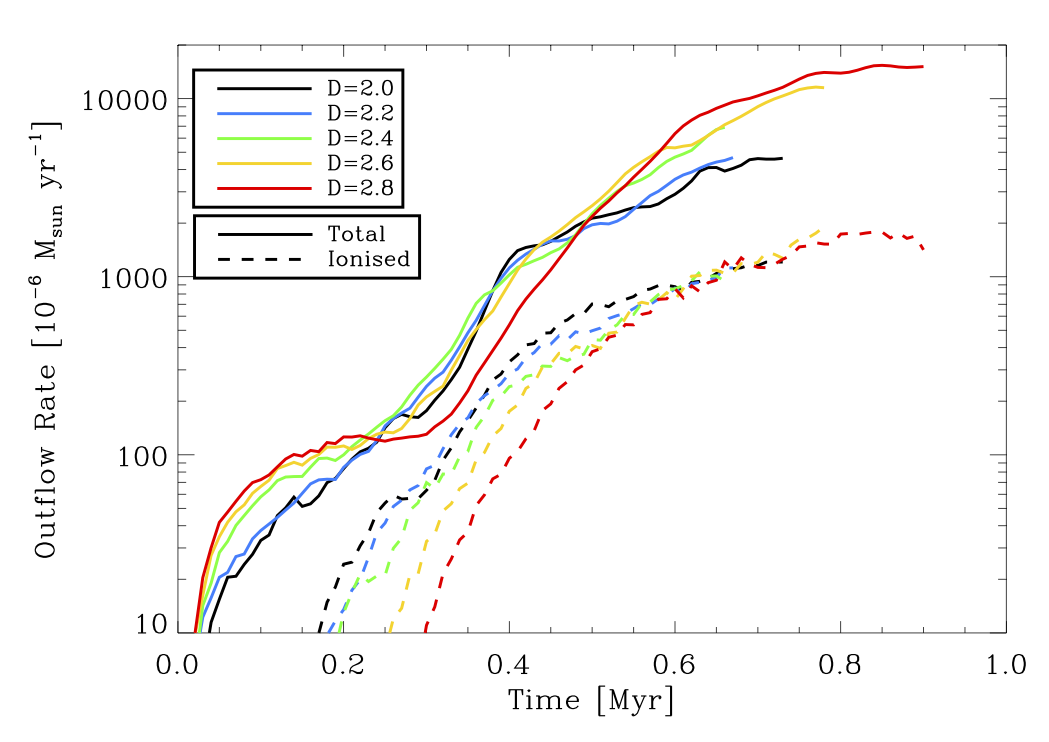} 
\caption{Mass outflow rates through the initial surface of the sphere at radius $6.4\,{\rm pc}$ for $\mathcal{D}=2.0,\,2.2,\,2.4,\,2.6\;{\rm and}\;2.8$. Solid lines show the total outflow rate, whereas dashed lines show the amount of outflowing ionised gas.}
\label{FIG_MASSLOSS}
\end{center}
\end{figure}

Fig. \ref{FIG_MASSLOSS} shows the total rate at which mass flows out through the initial boundary of the cloud (at radius $6.4\,{\rm pc}$), as well as the rate for the ionised gas only, both as functions of time. By $1\,{\rm Myr}$, the total rate is $\dot{M}_{_{\rm TOT}}\sim 0.5\;{\rm to}\;1.5\,\times 10^4\,{\rm M}_\odot\,{\rm Myr}^{-1}$, and  $10\;{\rm to}\;25\%$ of this is ionised gas. The dependence of $\dot{M}$ on $\mathcal{D}$ is rather weak.

Fig. \ref{FIG_EKIN}a shows the root-mean-square velocity for all the gas and for the neutral gas only, plus the mean radial velocity of the neutral gas, all as functions of time, $t$, for the runs with an ionising star at the centre. By $\sim 1\,{\rm Myr}$, the radial velocity of the neutral gas is approaching the escape speed from the cloud surface, and the ionised gas is travelling outwards even faster. We conclude that the cloud will be dispersed by $\sim 1\;{\rm to}\;2\,{\rm Myr}$, (where the longer time corresponds to lower ${\cal D}$). Fig. \ref{FIG_EKIN}a also shows the root-mean-square velocity for all the gas when there is no ionising source. In this case the velocities are generated by self-gravitational acceleration, and are much smaller. Therefore we can affirm that it is the action of the ionising star that dominates the cloud dynamics.

\begin{figure}
\includegraphics[width=90mm]{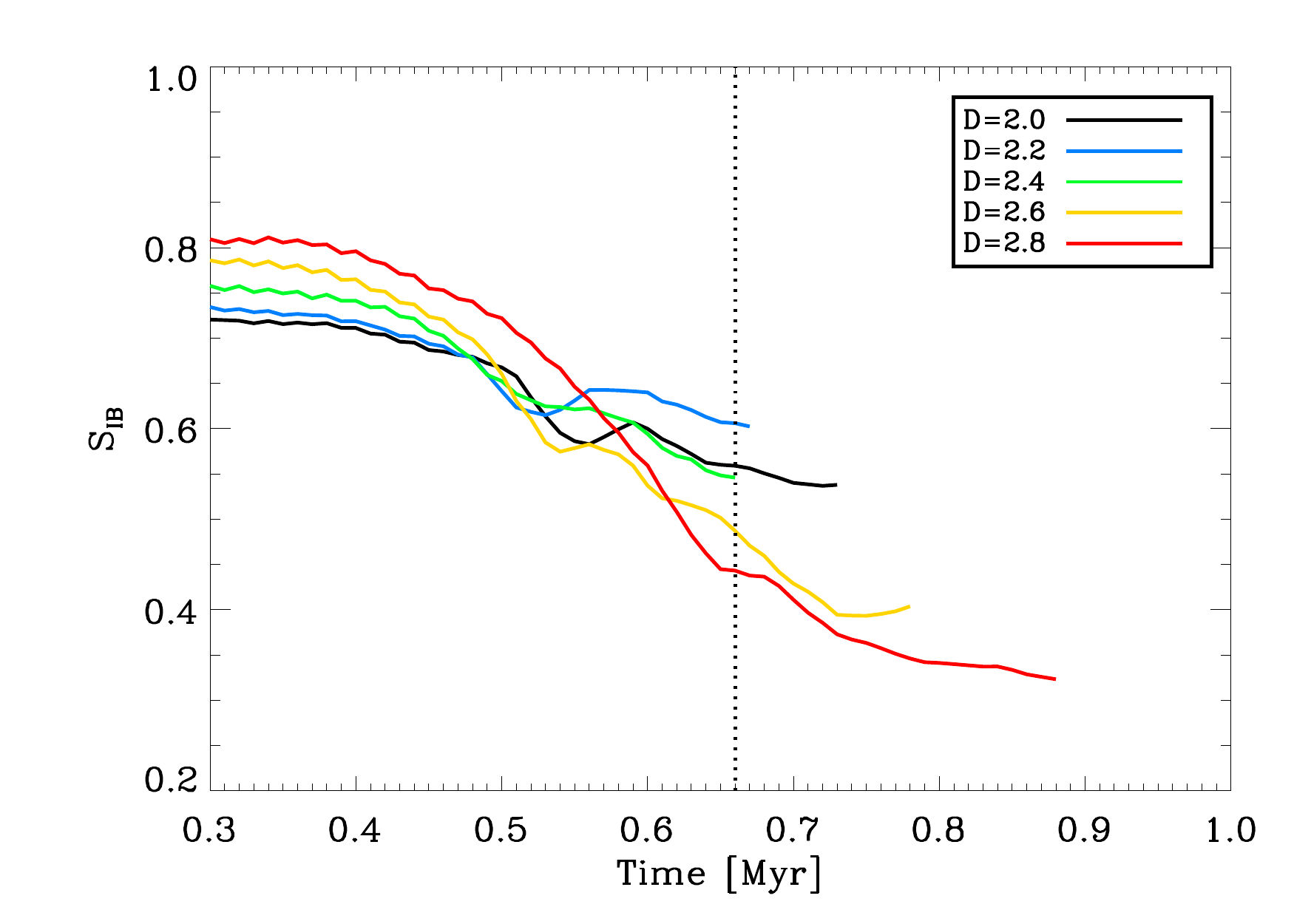} 
\caption{The fraction of the sky, as seen from the ionising star, that is ionisation-bounded, $f_{_{\rm IB}}$, as a function of time, $t$. From top to bottom, $\mathcal{D}=2.0,\,2.2,\,2.4,\,2.6\;{\rm and}\;2.8$.}
\label{FIG_SIB}
\end{figure}

\section{Turbulence driving \& density evolution}\label{sec5}%

\subsection{Driving turbulence}

The injection of bulk kinetic energy by expanding H{\sc ii} regions may make a significant contribution to the driving of turbulence in molecular clouds. \citet{Gritschneder2009} have shown that ionising radiation is effective in driving turbulence in the surrounding ISM; in particular they find that the fraction of turbulent energy contained in compressional modes is enhanced relative to the equipartition value of $1/3$. The simulations presented here admit an even more unequivocal opportunity to evaluate the role of ionising radiation in driving turbulence, since there are no initial turbulent velocities, and therefore we do not need to take account of dissipation. The total kinetic energy must be attributed to ionising feedback or to self-gravity. By performing comparison simulations with no ionising source, we can estimate accurately the contribution from ionising feedback.

\begin{figure*}
\begin{center}
\begin{tabular}{cc}
\includegraphics[width=90mm]{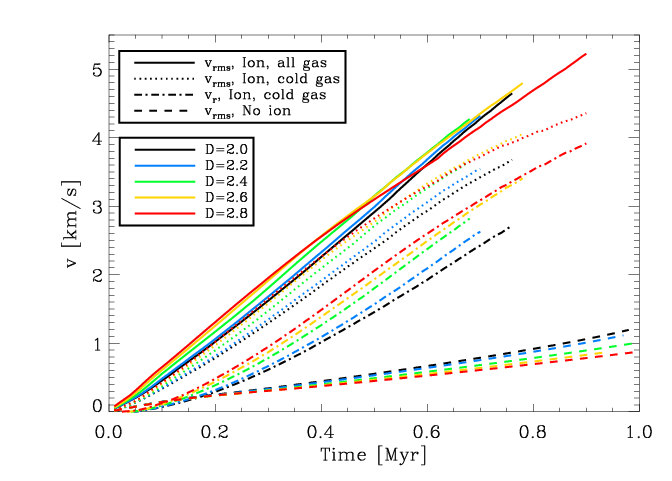} &
\includegraphics[width=87mm]{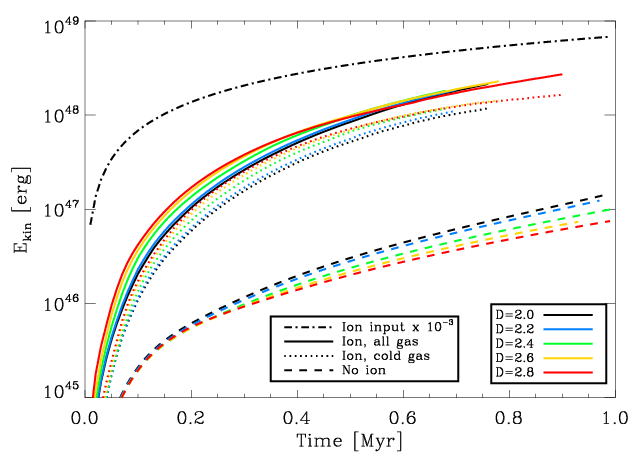} 
\end{tabular}
\caption{{\sc Left panel:} The full and dashed curves show the root-mean-squared velocity for all the gas in, respectively, the simulations with and without ionising feedback. The dotted and dot-dash curves show, respectively, the rms velocity and the radial velocity, for the neutral gas only in the simulation with ionising feedback. The thin horizontal line demarks the escape velocity from the surface of the initial cloud. {\sc Right panel:} The full and dashed curves show the total kinetic energy for all the gas in, respectively, the simulations with and without ionising feedback. The dash-dot line shows the total energy input in the form of ionising radiation, scaled down by $10^{-3}$.}
\label{FIG_EKIN}
\end{center}
\end{figure*}

From Fig. \ref{FIG_EKIN}a we see that, to a good approximation, the root mean square velocity increases linearly with time, both with (full curves) and without (dashed curves) ionising feedback. However, the rate without ionising feedback, $\sim 1\,{\rm km}\,{\rm s}^{-1}\,{\rm Myr}^{-1}$, is much smaller than the rate with ionising feedback, $\sim 6\,{\rm km}\,{\rm s}^{-1}\,{\rm Myr}^{-1}$. The corresponding rates of injection of kinetic energy, expressed per hydrogen nucleus, are $\dot{u}_{_{\rm KIN}}\sim 4\times 10^{-28}\,{\rm erg}\,{\rm s}^{-1}\,{\rm H}^{-1}$ without ionising feedback, and $\dot{u}_{_{\rm KIN}}\sim 1.4\times 10^{-26}\,{\rm erg}\,{\rm s}^{-1}\,{\rm H}^{-1}$ with ionising feedback; for comparison, the heating rate due to ionisation is $\dot{u}_{_{\rm ION}}\sim 5\times 10^{-22}\,{\rm erg}\,{\rm s}^{-1}\,{\rm H}^{-1}\,(\rho/6.2\times 10^{-22}\,{\rm g}\,{\rm cm}^{-3})$ -- where we have normalised the density to the mean value in the initial cloud.

Fig. \ref{FIG_EKIN}a also shows the root mean square velocity (dotted curves) and the radial velocity (dash-dot curves) for the neutral gas only, in the case where there is ionising feedback. We see that even the neutral gas acquires a large amount of bulk kinetic energy; and that a large part of this is invested in ordered expansion. The radial velocity is approaching $4\,{\rm km\;s}^{-1}$ at the end of the simulation, which is comparable to the bubble expansion velocity found in IC 1396, a shell-like HII region with a comparable radius \citep{Patel1995}. However, even if we discount the part invested in ordered expansion, the remaining random kinetic energy ("turbulence") is injected into the neutral gas at a greater rate by ionising feedback ($\sim 2 - 4\;{\rm km\;s}^{-1}$), than by gravity when there is no ionising feedback. Moreover, after $\sim 1{\rm Myr}$ the mean radial velocity of the neutral gas is approaching the escape velocity from the initial surface of the cloud ($3.7\,{\rm km}\,{\rm s}^{-1}$), and therefore the cloud is being effectively dispersed.

We note that some of the kinetic energy invested in ordered expansion could be transformed into turbulent energy when the expanding shell interacts with the surrounding interstellar medium. For instance, in their three-dimensional radiation-hydrodynamics simulations, studying the ionisation of turbulent clouds without self-gravity, \citet{Mellema2006} derive root-mean square velocities for the cold gas. In particular, they find that, after the initial $\sim 10^5$ yr, the difference between the one-dimensional rms velocity of the cold gas and the average expansion velocity of the cold gas is roughly constant at $\sim 4\,{\rm km}\,{\rm s}^{-1}$. Thus, the pre-existing turbulence of $2 \;{\rm km\;s}^{-1}$ seems to be maintained and enhanced by the ionising feedback. Therefore, the amount of energy, which is found to be transformed into random motions is comparable to our simulations.

Fig. \ref{FIG_EKIN}b shows the evolution of the bulk kinetic energy, both with (full curves) and without (dashed curves) ionising feedback; also plotted is the net energy input in the form of ionising radiation (dotted curve, scaled down by $10^{-3}$). By $\sim 1\,{\rm Myr}$,  the total radiative energy injected by the ionising source is $7\times 10^{51}\,{\rm erg}$. A part of the injected energy is transformed into kinetic energy $\sim 3\times 10^{48}\,{\rm erg}$. In the run without ionisation, only gravitational potential energy may be transformed into kinetic energy. We find that this results in $\sim 10^{47}\,{\rm erg}$. Thus ionising feedback is much more effective than gravity at generating bulk kinetic energy, at least on this timescale ($\sim 1\,{\rm Myr}$), but still very inefficient ($\sim 0.05\%$). Similar results have been found by \citet{Dale2005}, who model somewhat smaller, less massive clouds, having comparable mean densities; they do not place an ionising source at the center, but instead evolve a molecular cloud with turbulent initial conditions to form a massive star self-consistently. In radiation-magneto-hydrodynamic simulations \citet{Gendelev2012} have recently shown that the inclusion of a magnetic field may enhance the input of kinetic energy to the surrounding clouds, especially if the ionising star forms at the edge of a molecular cloud and generates a blister-type H{\sc ii} region. However, the energy only increases by a factor of $\sim 2$ in this case, so the efficiency is still $\la 0.1\%$.

\subsection{Timescales for spontaneous star formation}

Since no star formation occurs in the comparison simulations during the first $\sim 0.9\,{\rm Myr}$, we can infer that, up to this point, all star formation in the simulations with ionising feedback is triggered. To understand this result, consider the free-fall time in gas with density $\rho$,
\begin{eqnarray}\nonumber
t_{_{\rm FF}}&=&\left(\frac{3\,\pi}{32\,G\,\rho}\right)^{1/2}\\\label{freefalltime}
&=&2\,{\rm Myr}\,\left(\frac{\rho}{10^{-21}\,{\rm g}\,{\rm cm}^{-3}}\right)^{-1/2}\,.
\end{eqnarray}
and the sound crossing time for a cloud of neutral gas with radius $R$,
\begin{eqnarray}
t_{_{\rm SC,N}}&=&\frac{R}{c_{_{\rm N}}}\\
&=&3\,{\rm Myr}\,\left(\frac{R}{\rm pc}\right)\,.
\end{eqnarray}
Spontaneous (i.e. untriggered) star formation cannot occur before $0.9\,{\rm Myr}$, unless there are coherent lumps of gas with $t_{_{\rm FF}}<0.9\,{\rm Myr}$ and $t_{_{\rm FF}}\ll t_{_{\rm SC,N}}$. The first condition requires lumps with density $\rho >5\times 10^{-21}\,{\rm g}\,{\rm cm}^{-3}$, and the second condition requires coherent lumps with radius $R\gg 0.3\,{\rm pc}\,(\rho/5\times 10^{-21}\,{\rm g}\,{\rm cm}^{-3})^{-1/2}$. From inspection of the density PDFs in Fig. \ref{FIG:RHOPDF_PDF_ini}, it is clear that there is very little gas at such high densities, and what there is is unlikely to be in sufficiently large coherent lumps.

\begin{figure*}
\begin{center}
\begin{tabular}{c c}
$\mathcal{D}=2.0$ without ionisation & $\mathcal{D}=2.0$ with ionisation \\
\includegraphics[width=90mm, angle=0]{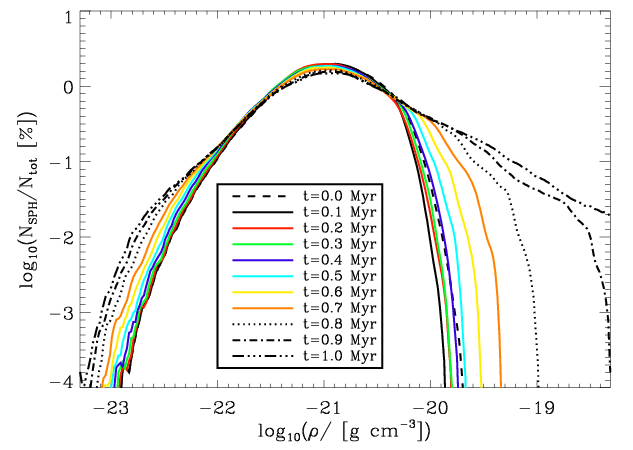} &
\includegraphics[width=95mm, angle=0]{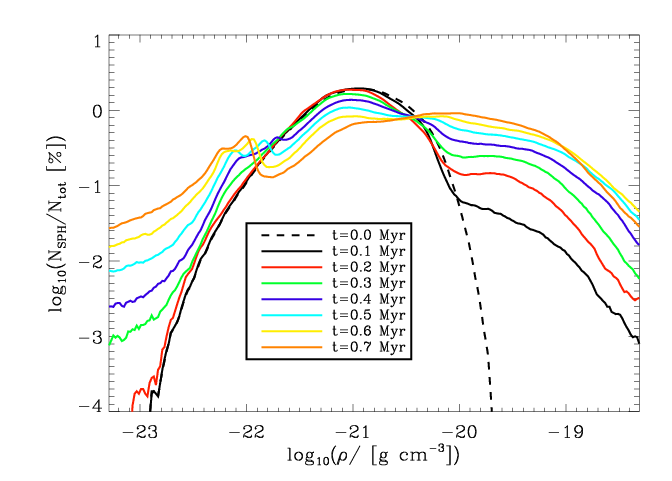} \\
$\mathcal{D}=2.8$ without ionisation & $\mathcal{D}=2.8$ with ionisation \\
\includegraphics[width=90mm, angle=0]{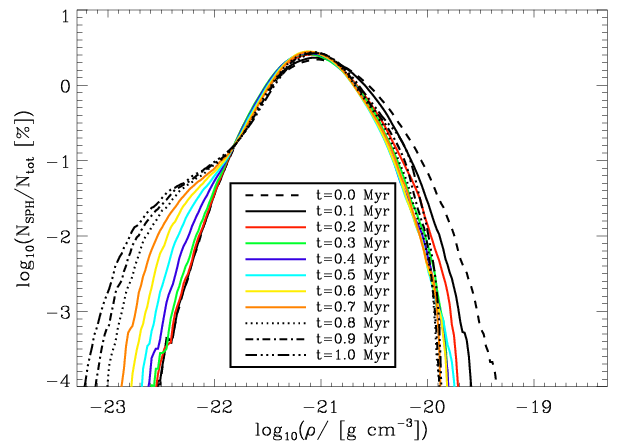} &
\includegraphics[width=90mm, angle=0]{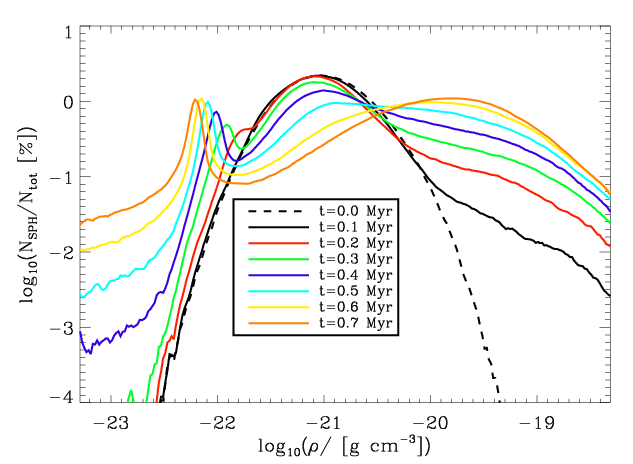} 
\end{tabular}
\caption{Time evolution of the density PDFs. The left panels are without ionising feedback, and the right panels are with ionising feedback. The top row is for ${\cal D}\!=\!2.0$, and the bottom row is for ${\cal D}\!=\!2.8$.}
\label{FIG:RHOPDF_PDF_evo}
\end{center}
\end{figure*}

\begin{figure*}
\begin{tabular}{c c}
\includegraphics[width=90mm]{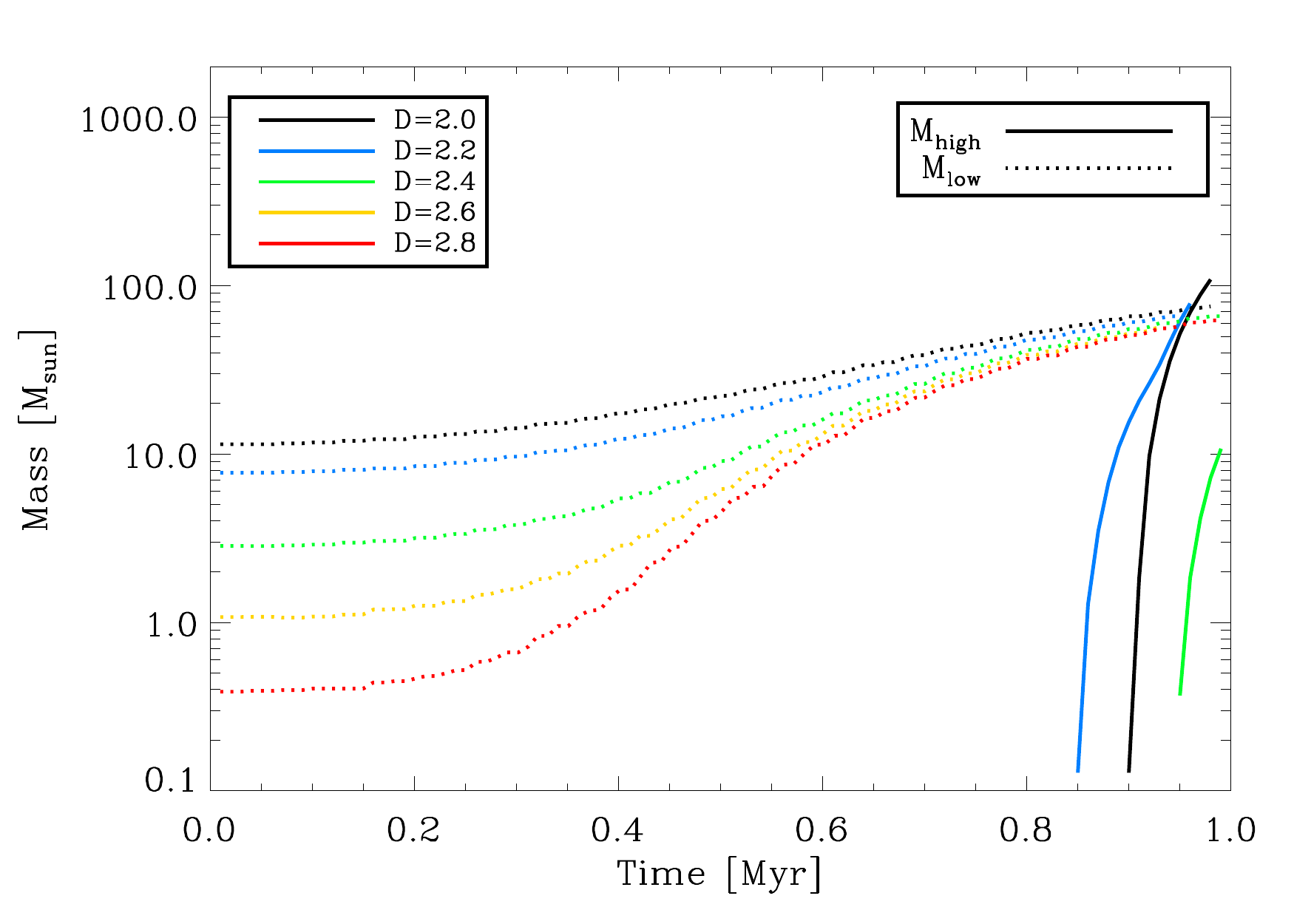} 
\includegraphics[width=90mm]{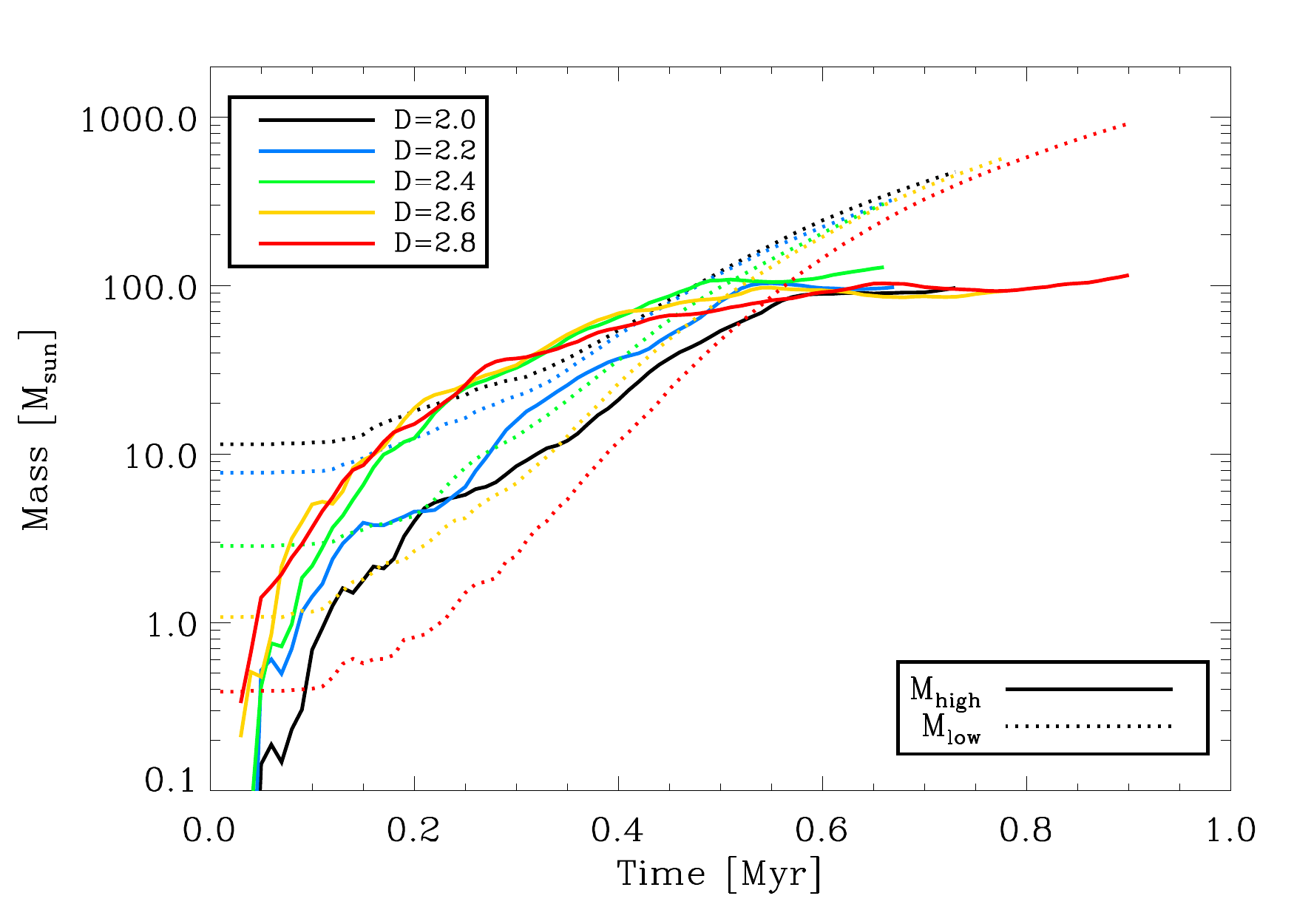} &
\end{tabular}
\caption{The full lines show the mass at high density ($\rho> 6\times 10^{-19}\,{\rm g}\,{\rm cm}^{-3}$), and the dotted lines show the mass at low density ($\rho<10^{-22}\,{\rm g}\,{\rm cm}^{-3}$), as a function of time, for $\mathcal{D}=2.0,\,2.2,\,2.4,\,2.6\;{\rm and}\;2.8$. The left panel is without ionising feedback, and the right panel is with ionising feedback.}
\label{FIG_MHIGHLOW}
\end{figure*}

\subsection{Density evolution}

Density PDFs can be used to characterise the structure of molecular clouds \citep[e.g.][]{Padoan2002}. Fig.\ref{FIG:RHOPDF_PDF_evo} demonstrates the influence of ionising feedback on the density PDFs of the molecular clouds with $\mathcal{D}=2.0$ and $\mathcal{D}=2.8$.

Without ionising feedback the evolution is very moderate, and simply involves a gradual broadening of the PDF, until the time approaches $\sim 0.9\,{\rm Myr}$. At this juncture a high-density tail suddenly develops, for ${\cal D}=2.0$, signalling the onset of collapse and spontaneous star formation. This is in good agreement with recent observational \citep{Kainulainen2009, Kainulainen2011, Lombardi2010} and theoretical \citep{Kritsuk2011, Girichidis2012} findings.

With ionising feedback, the evolution of the density PDF is much more rapid and distinctive. A narrow low-density peak develops  at densities $\rho\la 10^{-22}\,{\rm g}\,{\rm cm}^{-3}$, representing an expansion wave driven by hot ionised gas escaping into the surrounding ISM. At the same time a much broader high-density peak forms  at densities $\rho\ga 3\times 10^{-21}\,{\rm g}\,{\rm cm}^{-3}$, corresponding to neutral gas compressed by the overpressure of the H{\sc ii} region; this neutral gas is the material for triggered star formation. 

\subsection{Mass evolution}\label{massevo}

Fig. \ref{FIG_MHIGHLOW} shows the total mass in very low-density gas, $M_{_{\rm LOW}}$, and in very high-density gas, $M_{_{\rm HIGH}}$, as a function of time, for $\mathcal{D}=2.0,\,2.2,\,2.4,\,2.6\;{\rm and}\;2.8$. Here very low density means the escaping ionised gas in the expansion wave with $\rho<10^{-22}\,{\rm g}\,{\rm cm}^{-3}$, and very high density means gas which is sufficiently dense to couple thermally to the dust, $\rho>6\times 10^{-19}\,{\rm g}\,{\rm cm}^{-3}$, since this gas is very likely to condense into new stars.

Without ionising feedback, the mass in low-density gas increases with time as the density substructure relaxes, especially for clouds with high $\mathcal{D}$. After $\sim 1\,{\rm Myr}$, $M_{_{\rm LOW}}$ approaches $100\,{\rm M}_\odot$ for all ${\cal D}$. $M_{_{\rm HIGH}}$ is negligible in all runs until $t\ga 0.9\,{\rm Myr}$, when, for low ${\cal D}\la 2.4$, it increases rapidly, signalling the onset of spontaneous star formation. 

With ionising feedback $M_{_{\rm LOW}}$ rises even more rapidly than without ionising feedback, as gas is blown out into the surrounding ISM, and the cloud is dispersed. By $t=0.66$ Myr, $M_{_{\rm LOW}}$ has already reached $500\,{\rm M}_\odot$, and  by $t>0.85\,{\rm Myr}$, $M_{_{\rm LOW}}$ has passed $10^3\,{\rm M}_\odot$. At the same time, $M_{_{\rm HIGH}}$ increases rapidly up to $\sim 0.4\,{\rm Myr}$, and then saturates at $\sim 100\,{\rm M}_\odot$, for all ${\cal D}$. $M_{_{\rm HIGH}}$ is the mass that is available to form stars during the next $\sim 0.1\,{\rm Myr}$; note that it does not include the mass that is already in sinks.

Thus, although ionising feedback triggers star formation, it also disperses the cloud on a comparable timescale. Consequently, triggered star formation only accounts for a few percent of the cloud's mass. Any other star formation must occur spontaneously, promptly after the ionising star switches on.

\section{Conclusions}\label{sec6}%

We have used high-resolution 3D SPH simulations to explore the effect of a single O7 star emitting photons at $10^{49}\,{\rm s}^{-1}$ and located at the centre of a molecular cloud with mass $10^4\,{\rm M}_{_\odot}$ and radius $6.4\,{\rm pc}$. We focus on the shell structure and the dynamical impact of the ionising radiation on the surrounding molecular cloud as a function of its fractal dimension,  $\mathcal{D}$. We find that, although the shell morphology is strongly dependent on $\mathcal{D}$, global parameters like the total outflow rate, the mass in high and low density gas, the injected kinetic energy, and the average bubble radius are in general almost independent of $\mathcal{D}$.

Concerning the structures of the evolving H{\sc ii} regions, there is a morphological transition as $\mathcal{D}$ is increased from low to high values. Clouds with low $\mathcal{D}\la 2.2$ are dominated by extended density structures. Consequently, when the ionising star switches on they develop (i) large holes, through which ionised gas escapes into interstellar space, and (ii) between the holes, extended, rather smooth, shell-like structures where the ionisation front is bound by high density gas. We call these clouds {\it shell-dominated}. Conversely, clouds with high $\mathcal{D}\ga 2.6$ are dominated by compact density structures. Consequently, when the ionising star switches on, they develop (i) a rather uniform network of small holes through which the ionised gas can escape into the surrounding interstellar medium, and (ii) between these small holes there are numerous pillar-like structures protruding into the H{\sc ii} region. Therefore, we call these clouds {\it pillar-dominated}. The transition between shell-dominated and pillar-dominated H{\sc ii} regions occurs for clouds with fractal dimension $\mathcal{D}\approx 2.4$, which is close to the mean observed fractal dimension in Galactic molecular clouds. The H{\sc ii} region forming in a cloud with $\mathcal{D}=2.4$ nicely resembles observed H{\sc ii} regions \citep{Walch2011}. We note that the presence of pillars is a sign of there having been small scale density structures in the parental molecular cloud; these might, for example, have been generated by turbulence.

We discuss the influence of ionising feedback on the kinetic energy budget, and on the density structure of the clouds. Overall, the conversion of the energy of ionising radiation into kinetic energy is extremely inefficient, $\la 0.1\%$. However, on the short timescale before the cloud is dispersed, $\sim 1\;{\rm to}\;2\,{\rm Myr}$ with $\dot{M}_{_{\rm TOT}} \sim 10^{-2} {\rm M}_\odot {\rm yr}^{-1}$, ionising feedback is much more effective at injecting kinetic energy than self-gravity; in comparison, for simulations without ionising feedback, the amount of kinetic energy injected after $\sim 1\,{\rm Myr}$ is 40 times smaller. The kinetic energy injected by ionising feedback should not be termed turbulent, since much of it is invested in the ordered expansion that disperses the cloud.

Thus, ionising feedback is very effective at triggering star formation, in the sense of accelerating the rate of star formation, but it also terminates star formation by blowing the cloud apart, on a timescale $\sim 1\;{\rm to}\;2\,{\rm Myr}$. Consequently, triggered star formation is rather short-lived, and only converts a few percent of the cloud's mass into stars, thus limiting the total star formation efficiency of a molecular cloud. Any other young stars formed in the cloud must therefore have been formed before or promptly after the ionising feedback switched on.


\section*{Acknowledgments}
We thank Dr. Thorsten Naab for valuable discussion of the manuscript. SW acknowledges support by the Marie Curie RTN \textsc{CONSTELLATION} and the DFG Priority Programme No. 1573. The simulations have been performed on the Cardiff Arcca Cluster. APW gratefully acknowledges the support of a rolling grant from STFC. The work of TGB was funded by STFC grant ST\/H001794\/1. RW acknowledges support from the Czech Science Foundation grant 209\/12\/1795. DAH is funded by a Leverhulme Trust Research Project Grant (F/00 118/BJ).
\bibliographystyle{mn2e}
\bibliography{references}

\clearpage

\end{document}